\documentclass[%
 reprint,
 amsmath,amssymb,
 aps,
]{revtex4-2}

\usepackage{graphicx}% Include figure files
\usepackage{subcaption}
\usepackage{dcolumn}% Align table columns on decimal point
\usepackage{tabularx}
\usepackage{multirow}
\usepackage{bm}% bold math
\usepackage{amsfonts,color} % Color text for comments
\usepackage[table,xcdraw]{xcolor}
\usepackage[english]{babel}
\usepackage{lipsum} 
\usepackage{array}

\definecolor{Gray}{gray}{0.9}

\newcolumntype{Y}{>{\centering\arraybackslash}X}
\newcolumntype{Z}{>{\raggedright\arraybackslash}X}

\newcolumntype{P}[1]{>{\raggedright\arraybackslash}p{#1}}

\begin{document}

\preprint{APS/123-QED}

\title{
Physics PhD student perspectives on the importance and difficulty of finding a research group
}

\author{Mike Verostek}
  \email{mveroste@ur.rochester.edu}
\affiliation{
 Department of Physics and Astronomy, University of Rochester, Rochester, New York 14627 
}
 \affiliation{School of Physics and Astronomy, Rochester Institute of Technology, Rochester, New York 14623}

 \author{Casey W. Miller}
 \affiliation{School of Chemistry and Materials Science, Rochester Institute of Technology, Rochester, New York 14623}

 \author{Benjamin M. Zwickl}
 \affiliation{School of Physics and Astronomy, Rochester Institute of Technology, Rochester, New York 14623}

\date{\today}

\begin{abstract}

Joining a research group is one of the most important events on a graduate student's path to becoming an independent physics researcher and earning a PhD.  However, graduate students' perspectives on the experience of finding a research group are not well-documented in the literature.  Understanding these perspectives is crucial for evaluating whether departments are providing students with adequate support while they search for a research group, and how difficulties during this process contribute to attrition.  Semi-structured interviews with $N=20$ first and second year physics PhD students reveal that incoming graduate students see joining a research group as a significant decision, and recognize that it may impact whether they will be able to complete the program.  We found that students who struggled to find a group felt isolated and worried about falling behind their peers, whereas students who were able to immerse themselves in a positive group environment reported increased sense of belonging in their programs.  The process of finding a research group often held differential importance for students identifying as women and non-binary, who at times reported having to deprioritize their preferred research topic in order to be part of a more inclusive working environment.  Although incoming graduate students characterized joining a research group as a significant decision, they often felt unprepared to make it.  Moreover, they perceived an overall lack of guidance and structure from their departments, and characterized coursework as a barrier to searching for a group.  Our findings suggest that providing students with better support during their group search process could help improve retention, particularly for traditionally underrepresented students, and improve students' overall satisfaction in their graduate programs.

%\begin{description}
%You may use the \texttt{description} environment to structure %your abstract;
%use the optional argument of the \verb+\item+ command to give %the category of each item. 
%\end{description}

\end{abstract}

\maketitle
%\tableofcontents

\section{\label{sec:Introduction}Introduction and Background}

Increasing the retention rate of physics graduate students is of paramount importance for both students and departments.  Current data indicates that the retention rate of physics PhD students is approximately 50\%, with attrition from PhD programs disproportionately affecting traditionally underrepresented students in physics \cite{cgs2008phd, cgs2015minorityphd, lott2009doctoral, miller2019typical}.  For students, attending graduate school requires significant personal and financial sacrifices as they pursue long-held career plans and aspirations.  Leaving a PhD program often means grappling with feelings of failure and disappointment, which can have major adverse effects on students' future mental health and financial well-being \cite{lovitts2002leaving}.  Graduate programs are not left unscathed by these outcomes either.  Departments expect their investment in students to yield productive research for the institution, but early departures cannot contribute to this endeavor.  Moreover, departments must invest future resources into recruiting and supporting new students \cite{lovitts2002leaving, golde2005role}.  

Previous studies on graduate attrition across STEM and non-STEM disciplines indicate that a negative advising relationship is one key factor that motivates students to leave \cite{lovitts2002leaving, devos2017doctoral, bair2004doctoral, rigler2017agency, golde1998beginning, jacks1983abcs}. As \citeauthor{lee2008doctoral} succinctly notes in her multidisciplinary study on doctoral advising, it is widely acknowledged in the academic community that an advisor ``can make or break a PhD student'' \cite{lee2008doctoral}.   Yet PhD students often find navigating their first year of study to be difficult \cite{gardner2007heard}, and if the process of finding a group is difficult for students to navigate, they may be less likely to find a group that provides them with a fulfilling research experience.  Thus, the process by which students find a research group may be an important underlying cause of student attrition.  Physics remains one of the least diverse fields across STEM disciplines \cite{noauthor_nsfgov_nodate}; however, as the number of physics PhDs granted across the US grows \cite{aip2021trends} and significant resources are directed toward diversifying the field \cite{aps_idea, aip2020teamup}, investigating the underlying factors behind high attrition is imperative for assuring that PhD students of all backgrounds are given the support they need to thrive \cite{sachmpazidi2021departmental}.  Pursuant to this goal, this paper investigates how physics PhD students experience and characterize their search for a research group and advisor.  

Advisors are largely responsible for developing students into independent physics researchers, providing a positive and fulfilling graduate experience, and ultimately shepherding advisees through the PhD process.  Additionally, choosing an advisor affords students access to a community of postdocs and graduate students in their lab, which can be equally critical for professional development.  At both the undergraduate and graduate levels, becoming a part of a research group can give STEM students opportunities to develop their identities as scientists \cite{malone2009narrations, graham2013increasing} while increasing their sense of belonging to their research community \cite{o2017sense, stachl2020sense,alaee2022impact, dolan2009toward, lopatto2007undergraduate}, in addition to expanding their research knowledge and skills \cite{juliano2001critical, alaee2022impact, lopatto2004survey, hunter2007becoming}.  Across STEM fields, high-quality advising relationships are known to support a number of positive student outcomes  \cite{national2020science, aikens2017race, byars2015culturally, lisberg2018mentorship}, including successful doctoral program completion \cite{devos2017doctoral, bair2004doctoral}. Hence, much research has focused on identifying the characteristics indicative of productive mentorship \cite{schlosser2003qualitative, schlosser2011multiculturally, barnes2010characteristics, bargar1983advisor}.  However, these studies were multidisciplinary and sometimes included non-STEM students, and they largely focus on advisor-advisee relationships \textit{after} they have formed. Comparatively less research has explored how those relationships came to exist in the first place.

Several multidisciplinary studies have investigated which factors PhD students most highly prioritize while looking for a group, finding that funding availability and research interest tended to be most important \cite{zhao2007more, golde2001cross, joy2015doctoral}.  Still, little research has sought to answer questions surrounding students' attitudes and experiences during the group search process, how they gather information about prospective groups, and how they make sense of that information.  Notable exceptions include a pair of longitudinal qualitative studies by \citeauthor{maher2019doctoral}, who detailed how biology PhD students leverage a combination of formal lab rotations and informal communication with peers to search for a group that both matches their research interests and has a positive social environment \cite{maher2019doctoral, maher2020finding}.  These studies specifically focused on how the structure imposed by lab rotations influenced students' experiences, finding that students' research interests were not well-formed by the time they arrived in graduate school and were strongly shaped by their rotations. They also described how availability of funding could limit students' agency in choosing a group. 

No prior research has investigated the group search phenomenon in the context of physics graduate programs.  The process by which physics graduate students go about finding research groups is therefore relatively unexplored, and the applicability of existing research to physics graduate education is unclear due to the stark differences in how STEM doctoral programs facilitate finding a research group \cite{artiles2023doctoral}.  For instance, biology departments and medical programs typically employ highly formalized rotation systems, as detailed by \citeauthor{maher2019doctoral}  Such formal requirements are comparatively rare in physics graduate programs \cite{feder2020phd}, where the most common structures through which students learn about prospective groups are faculty research seminars \cite{artiles2023doctoral}.  These varying disciplinary rules and norms uniquely shape students' experiences \cite{gardner2010contrasting, golde2005role}, and signal the importance of studying the group search process within specific disciplinary contexts.      

To begin addressing this gap, we interviewed first and second year physics PhD students to better understand their perspectives on navigating process of finding a research group.  In particular, we sought to gain insight into how graduate students characterize the importance of finding a research group, and how well-supported they feel throughout their search.  Although faculty and senior graduate students often cite choosing a research group as one of the most crucial decisions doctoral students must make \cite{bloom1999ph, barres2013pick, jabre2021ten}, it is unclear how incoming students think about this decision.  The impact that the search itself may have on students if they encounter difficulties joining their preferred group (or any group at all) is also uncertain.  By exploring these questions, we uncover several ways that the group search process affects overall graduate student satisfaction in their program, as well as illustrate how this understudied aspect of doctoral education may play an important role in retention and leaving.

Our study was guided by the following research questions:    

\begin{enumerate}
    \item How do physics PhD students characterize the significance of finding a research group and advisor in graduate school?
    \item In what ways does the process of searching for a research group and advisor impact students' overall graduate experience?
    \item How do students describe their ability to navigate the process of searching for a research group?
    \item What factors contribute to these descriptions?
\end{enumerate}

\begin{table*}[t]
\centering
\def\arraystretch{1.1}%  1 is the default
\begin{tabularx}{\textwidth}{|X|Y|Y|Y|Y|Y|Y|}
\cline{1-7}
 &White/caucasian & Hispanic, Latinx, or Spanish origin& Black or African American & Asian & From multiple races & Total \\ \hline
\multicolumn{1}{|l|}{Male} & 3 & 3 & 0 & 1 & 0 & 7 \\ \hline
\multicolumn{1}{|l|}{Female} & 4 & 2 & 2 & 2 & 1 & 11 \\ \hline
\multicolumn{1}{|l|}{Non-binary} & 1 & 1 & 0 & 0 & 0 & 2 \\ \hline
\multicolumn{1}{|l|}{Total} & 8 & 6 & 2 & 3 & 1 & 20 \\ \hline
\end{tabularx}
\caption{\label{tab:demographics1} Demographic breakdown of the data used in this analysis.  Demographic information was gathered using a fixed-choice Qualtrics survey prior to each interview.}
\end{table*}

\section{\label{sec:Method}Method}

This study is part of a larger analysis aimed at systematically characterizing the process by which PhD physics students search for a research group.  This overarching goal informed our interview protocol, which was inspired by cognitive task analysis (CTA) methods \cite{crandall2006working, clark2008early} and Dervin's sense-making method \cite{dervin1986neutral, dervin2003sensemaking}.  These methodologies are designed to elicit detailed descriptions of interviewees' thoughts and actions as they recount how they progressed toward a goal.  CTA and sense-making focus on understanding specific life experiences in great depth, often resulting in a rich data set.  

Our protocol asked students to construct a timeline of steps they took while searching for a research group.  We then asked students about any major questions and concerns they had at each step of their timeline, which often yielded comments about the importance of finding a group, general difficulties that students experienced, as well as concerns about their future in graduate school.  The analysis presented here primarily focuses on several of these major questions and concerns, while future work will leverage other aspects of the protocol.  The full interview protocol, as well as more detailed description of how this project fits into our larger analysis, is available in the Supplemental Material \cite{supplementalMats}.

% Our protocol is broadly broken into three stages.  First, we gathered a timeline of \textit{Steps} that students took in their search for a research group.  Students defined the start and end points of their stories, so these boundaries varied based on each student's individual experience.  Some students started their timelines during an undergraduate research experience.  Others were prompted to start thinking about advisors when they applied to graduate school, and some did not give serious thought to their research group until after they began the first year of their PhD program.  Timelines typically ended with students joining a group or continuing their search.  Examples of common steps on the timeline included ``Applying to graduate school'' or ``Attending visiting weekend.''  We then asked students to go through each step and identify their major \textit{Questions} and concerns regarding their search for a research group at that point in time.  This allowed us to gain insight into what information students were trying to find out, as well as which aspects of the process drove confusion and uncertainty. Lastly, we asked students to identify any sources of \textit{Help} that allowed them to resolve their question or concern, as well as any obstacles that \textit{Hurt} their ability to move forward.  This stage allowed us to understand the thoughts, actions, and events that helped or hurt students' ability to navigate their search for a group.  

Study participants were recruited by emailing graduate program directors and asking them to forward our recruitment letter to their first and second year graduate students.  We targeted these years of study because they were either in the process of or had recently completed searching for a research group.  We also intentionally chose to email programs of varying size and research activity to ensure a variety of institutional contexts were represented.  In total, we reached out to 18 graduate programs directly.  Ten programs forwarded our recruitment email to their first and second year graduate students.  Since we reached out to programs rather than individual students, we cannot know precisely how many students received invitations to participate.  However, physics graduate programs admit an average of 16 students per year \cite{aip2019firstyear}, meaning that the number of students invited to be interviewed was likely several hundred.  A \$25 Amazon gift card was offered as incentive for participation in our study.  

Since this project is conducted with the support of the Inclusive Graduate Education Network (IGEN), a partner of the American Physical Society (APS), one major goal of this work is to improve diversity and inclusion across physics graduate education.  The APS Bridge Program is a post-baccalaureate program designed to increasing the number of PhDs earned by underrepresented students in physics, and is currently one of the leading programs for diversifying graduate physics education \cite{aps_bridge_home}.  Hence, we also sent our recruitment information directly to current APS Bridge students to help ensure our data represented their experiences as well.

The sample of students in this analysis consists of 20 students representing 12 institutions; 5 students were from the Bridge program.  Based on a fixed-choice demographic survey administered prior to each interview via Qualtrics, $N=11$ interviewees identified as women, $N=7$ identified as men, and $N=2$ identified as non-binary or gender fluid.  $N=12$ were in their first year of physics graduate school while $N=8$ were in their second year.  $N=8$ identified as White/Caucasian, $N=6$ as Hispanic/Latinx or Spanish origin, $N=3$ as Asian, $N=2$ as Black or African American, and $N=1$ from multiple races.  $N=17$ students were from the US while $N=3$ were non-US students.  $N=12$ institutions were represented, varying in size and research activity.  Table \ref{tab:demographics1} offers a more detailed demographic breakdown.  Interviews were conducted over Zoom and took place from Fall 2022 to Spring 2023.  All names used throughout the paper are pseudonyms.  Other information, such as how far along each student was in their graduate program when the interview took place, as well as a more detailed breakdown of institution types and their program requirements, is available in the Supplemental Material \cite{supplementalMats}.

Once completed, interviews were transcribed and edited for grammar and clarity.  The transcripts became the subject of our thematic analysis, which followed the steps outlined in \citeauthor{braun2006using} \cite{braun2006using}. Analysis began with repeated and active reading of the data and generation of several emergent codes.  These codes were \textit{Significance}, \textit{General difficulties}, and \textit{Sense of belonging}.  \textit{Significance} was applied to statements about how choosing a research group will have or is already having a significant impact on students’ lives and careers. \textit{General difficulties} referred to statements about trouble navigating aspects of the group search process as a whole, especially comments wishing the department had provided more guidance.  \textit{Sense of belonging} was applied to statements about how joining a group helped students feel more integrated into the graduate program, or how being unable to find a group caused students to feel detached from the program and less able to succeed.  For each emergent code, we also applied an \textit{initial code}, which is a common first step in grounded theory approaches to data analysis.  Initial coding is an open-ended first cycle coding process that involves breaking down qualitative data into discrete parts, examining them closely, and applying codes that promote deep reflection on the contents of the data \cite{saldana2013introduction}.  Our initial codes typically consisted of a short word or phrase describing the excerpt and what we found important about it.

\begin{figure}[t]
\centering
\includegraphics[]{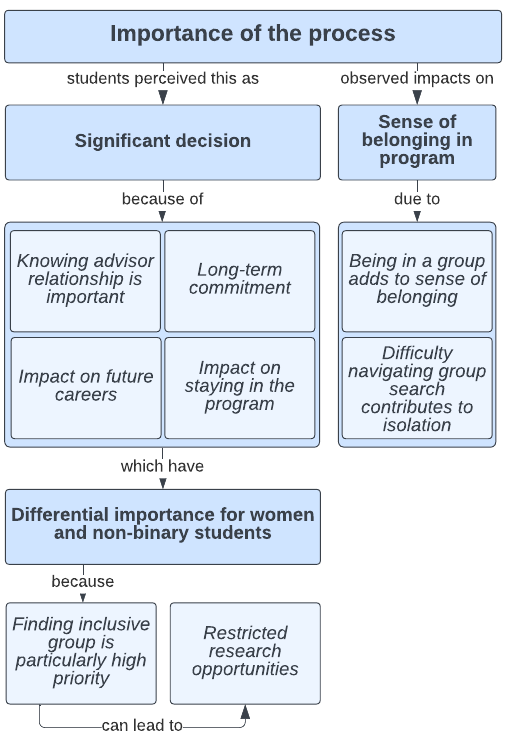}
\caption{\label{fig:importancethemes} A major theme that emerged was the overall importance of the group search process in students' doctoral experience. Students perceived choosing a group as a significant decision in their graduate careers.  Interviewees who identified as women or non-binary described finding an inclusive group as a particularly high priority, which we observed could restrict access to certain research opportunities.  We also observed that navigating the search process could have major impacts on students' sense of belonging in the program.}
\end{figure}

Codes were then sorted into themes and sub-themes.  The overarching theme of \textit{Importance of the process} is detailed in Section \ref{subsec:important} and summarized in Fig. \ref{fig:importancethemes}.  This theme describes the reasons we identified for why this process plays an important role in students' doctoral experience.  It is broken down into two sub-themes.  The first aligns with RQ1 and is called \textit{Significance of the decision}. This sub-theme characterizes why students believe finding a group is such a major decision, as well as why the decision holds differential importance for women and non-binary students. The second sub-theme aligns with RQ2 and is called \textit{Sense of belonging}.  This describes how the group search positively and negatively impacted students' belonging in their graduate programs.  

\begin{figure}[b]
\centering
\includegraphics[]{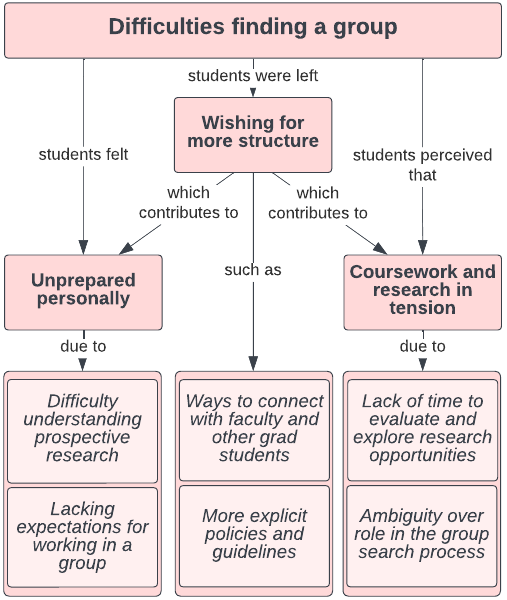}
\caption{\label{fig:difficultythemes} An overarching theme was that students experienced many difficulties while finding a group. Students sought more structure and guidance from departments, particularly regarding ways to connect with faculty and graduate students.  Improved guidance from faculty might have helped alleviate several other difficulties, including feeling unprepared for the search process and perceiving that their coursework was in tension with their ability to find a group.}
\end{figure}

Meanwhile, the overarching theme \textit{Difficulties finding a group} is detailed in Section \ref{subsec:difficulties} and summarized in Fig. \ref{fig:difficultythemes}.  It is broken down into three sub-themes.  The first aligns with RQ3 and describes how students feel \textit{Unprepared personally} to navigate the search process.  The other two, \textit{Wishing for more structure} and \textit{Coursework and research in tension} attend to RQ4 and illustrate several specific reasons for students' difficulties.

Themes were refined through discussion at weekly research meetings and a codebook was developed with final names and definitions for the themes.  Inter-rater reliability was done at this point to demonstrate validity of the codebook.  Drawing from the pool of excerpts that the author (MV) had coded, a selection of 40 random excerpts was given to a researcher unaffiliated with the project to code using the themes and sub-themes described above.  Agreement was over 90\%, and the few disagreements were primarily due to differing interpretations of an excerpt's meaning out of context rather than ambiguities in the codebook.  We also had the independent rater apply the codebook to a 30-minute section of a transcript.  Agreement here was also high; the rater matched all 17 cases in which the author (MV) applied a code to an excerpt, but also coded 2 extra excerpts as \textit{Sense of belonging}.  After discussion, we refined this code's inclusion criteria to more clearly demarcate when it should be applied.

\section{\label{sec:Results}Results}

\subsection{\label{subsec:important}Importance of the process}
\subsubsection{\label{subsubsec:importance}Students see joining a research group as an important decision} 

Students described finding a research group as a highly important part of their graduate school experience, characterizing it as a major commitment impacting their graduate careers and beyond. Four reasons were most commonly cited: 1) their individual relationship with the advisor is crucial, 2) joining a research group is a long-term commitment, 3) their choice of research group will impact their professional careers beyond the PhD, and 4) the perception that a poor working environment might lead them to leave their program.  Out of the 20 interviews we analyzed, 16 students explicitly cited one of these as a reason for why they believed finding the right research group was so significant.  Although the remaining four students did not make explicit comments about the importance of choosing a group, they demonstrated through their thoughts and actions that they invested a great deal of time into finding an advisor.

First, $N=7$ students described joining a group as a significant decision because they believed having a good individual relationship with their advisor was important to their success.  When evaluating potential groups, Olivia said that choosing the right advisor was ``the most important thing in grad school'' while Gabriela recalled that ``even before I started applying for graduate programs, all I heard was, `Your relationship with your PI can make or break your career. It's the most important relationship you'll have.'{''}  Carmen described feeling ``blessed'' because ``I've heard my fair share of horror stories with graduate students and advisor relationships, and I haven't had any of those experiences with my current advisor.''  These statements indicate students are aware that their individual advising relationships are crucial.  Students across the sample indicated that the multi-year duration of the advisor-advisee relationship was one major reason they wanted to ensure a positive individual relationship with their advisor. As expressed by Chloe, having a positive relationship with her advisor was important because ``they were going to be like, I don't wanna say `in charge of my life,' but kind of in charge of my life for the next five years, which was a little terrifying.''

Chloe's recognition that her advising relationship would last several years  alludes to the second reason that $N=9$ students cited for why finding a suitable research group was particularly significant: joining a research group meant making a long-term commitment to a particular research topic.  For instance, Matias felt apprehensive about dedicating years of his life to working on just a few projects: ``I was trying to figure out, like, which ones are projects that I can really get behind. Something that I can like, call it my own and do it for five years, you know? Which can get pretty hard. You're like, how are you going to make this decision now?''  These commitment concerns lingered for some students even after joining a group and working in it for several months.  Irene was enjoying her time working for a high energy experimental group the summer before starting graduate school, but officially committing to high energy research for the duration of her graduate career gave her pause.  She recalled, ``I wasn't like actively questioning it, but it was still like, could I do this for the next six years? Like, what? This is one project, but there are obviously other projects that would need to be done. So once this part's over, what would I do next? That I had to think about.''  Put more succinctly by Alex, who after working with a group over the summer and first semester of graduate school noted that ``Fear of commitment is always a thing. It's like, `Oh, what if I made a mistake?' Oh, dear.''  

The third reason students felt finding a group was so significant was the perception that their group choice would impact their careers after graduate school.  $N=5$ students had concerns about the impact that their choice would have beyond their PhD program.  Irene described feeling ``normal anxiety'' because she felt as though ``I'm committing to something for like the rest of my life. Oh, my gosh, that's terrifying.''  Matias expressed a similar concern, stating that ``There was also the fear, just like fear that by choosing a computational chemistry group, I was kind of securing, or kind of locking myself into a certain path. Like, what if biophysics was a better route for me?''  Another student, Dev, weighed his ability to gain employment in industry if he pursued research with a particular advisor: ``Basic science does not fit good job opportunities. I'm talking about quantum foundations. Maybe if I want to be in academia... But definitely not what industry is looking for.''  Thus, students acknowledged that their research groups would not only be important for persisting in graduate school, but would influence their careers beyond the PhD.

Lastly, part of the trepidation that students expressed when committing to a group was due to their knowledge that joining the wrong group can lead to a significantly more stressful PhD experience, and could result in leaving the graduate program altogether.  $N=4$ students, who all identified as women, voiced the possibility of needing to leave their programs due to a poor research group environment.  Rose, a second year grad student, recalled grappling with this concern. She said, ``I wanted to make sure I knew what was going on before I committed the next, you know, five to seven years of my life. And I'm not wanting to drop out. I saw those numbers as the president of [the Women in Physics Society]. I saw those numbers of people who attempted and didn't finish... And I was like, I'm gonna find someone here, and it's going to work out. How I'm gonna do that? I don't know. So that was a long process of figuring that out.''  Pauline shared similar concerns, stating, ``This is like six years in my life. And I don't want to be stuck in a kind of scenario where I feel like I have to drop out because I can't just take it in this group anymore.''  Another PhD student, Tabitha, had switched groups due to a poor working environment.  She framed her choice to change groups as a decision of whether to stay and ``get my PhD in three [more] years, and at the end of the three years be mentally exhausted, hate my research, hate academia, and be in a bad place mentally speaking'' or to leave and ``get my PhD in four years, five years, but be happy about my research and be in a good place mentally speaking.''  For these students, finding the right research group meant acknowledging that they were making a long-term commitment that could influence their degree completion and overall satisfaction with their graduate experience. 

%%%%%%%%%%%%%%%%%%%%%%%%%%%%%%%%%%%%%%%%%%%%%%

\subsubsection{\label{subsubsec:differential}Differential importance for women and non-binary physics students}  

The process of finding a research group often held differential importance for students in our sample who identified as women or non-binary. While most students sought to find a supportive work environment and recognized that a healthy lab culture would improve their graduate school experience, the criteria for what constitutes a `supportive group' was different for non-male students. Only women in our sample described the possibility that their research group could influence their persistence in the program, and whereas a majority of women (7 of 11) and all non-binary interviewees (2 of 2) described having to pare down their list of prospective research groups due to worries that the advisors would not foster inclusive environments, only one man (1 of 7) described a similar concern.  This disparity indicates that the present system for finding a research group inherently grants more research opportunities for male students, who did not feel the need to screen groups based on indicators of their inclusivity.  These nine non-male participants described looking for a number of these indicators both before and after admission to graduate school, including the diversity of the department's faculty, the diversity of individual groups' students, experiences of traditionally underrepresented students in a group, the advisor's outreach efforts, and their comfort level during one-on-one interactions with the advisor.

Selena said having other women in the department was ``really important'' to her, and wondered ``am I going somewhere where there are other women, where there are people of color, where I am going to stand a chance of seeing real mentorship from someone who might get me?''  She recalled that some schools had faculty whose research aligned with her interests, but she chose not to apply to those programs due to lack of diversity in the department : ``If I go through your department page, and all I see are white men, I'm not clicking any further... There's no way that there's not a single woman or person of color who was good enough for your faculty.''  Similarly, Gabriela characterized having a diverse research team as her ``biggest concern'' while evaluating groups.  For Gabriela, this requirement ``stemmed directly from my last lab'' where she was the only woman in a group of ten students.  Gabriela's experience  ``painted very clearly in my mind what I did not want the next five to six years to be,'' and expressed how she struggled to feel ``safe'' and ``comfortable'' working in her previous lab.  She emphasized, ``my biggest concern going forward is that I want to feel okay to take up space in lab.''  Finding a diverse and inclusive group was critical for these students, yet similar concerns were seldom voiced by male interviewees.    

The guidance students received regarding advisors' treatment of traditionally underrepresented students was particularly influential in shaping their opinion of prospective groups.  For example, Pauline described how ``I talked to one [graduate student] and I was like, he's not a woman, but a minority. And he seems to really like it there. And I was like, maybe it'll be fine then.''  For Pauline, the other student's status as a minority student in physics gave his opinion added credibility.  Similarly, Blake remembered being cautioned by other graduate students about certain advisors during visit weekend, saying ``There were a couple of people where I was warned by other grad students about like, this professor's a little verbally abusive towards his students, particularly towards women and gender minorities. You might not want to work for him.''  Heeding this advice, Blake chose not to pursue those groups.  Similarly, when Chloe met with a prospective group's only female graduate student, Chloe recalled, ``there were times where [the student] felt like she had been cut out of the writing process, or cut out of the process in general, and that sometimes it felt pretty gendered. And so especially when I had also gotten some weird vibes, that kind of helped me say, `Okay, you're not crazy. You're not just making this up.'{''}  Hence, both Chloe and Blake were driven away from research groups specifically because of concerns that their advisor would not treat them with respect due to their gender. 

More alarmingly, two of the 11 women interviewees reported that they were compelled to leave their groups after discovering that their advisors had been investigated for misconduct allegations related to their treatment of graduate students.  In one case, Tabitha was initially excited to work on her research project because it was the ``perfect'' topic, combining her interests in astronomy and computer science. Over the course of a semester of research though, she became disillusioned with the group due to the ``toxic'' working environment.  Still, she struggled with the decision to leave out of fear she would be damaging her career and prolonging her PhD.  Then, ``When I learned that my old advisor had open investigations because of sexual misconduct... that was a deal-breaker,'' and she left.  Tabitha opted to abandon the research she wanted to pursue in favor of a topic that ``wouldn't have been my first choice,'' but was under the supervision of a different advisor.   Similarly, when Pauline began working with her advisor, she characterized her research topic as something that she ``really wanted to pursue.''  She said that her advisor ``seemed like he was really interested in helping me grow as an individual and researcher. So I thought wow, I really hit the jackpot, like, perfect person, perfect PI.'' However, upon hearing from peers that her advisor had several ``serious allegations'' against him, she chose to reach out to one of his former graduate students.  ``I sent an email to one of his graduate students that worked there, like, three years ago. And she agreed to meet with me, we had a phone call. And really, she just said that if you're a woman, you shouldn't be in his group.''  Like Tabitha, Pauline left this group to pursue research that she found less interesting, but in an environment that she would ``be comfortable with for the next six years.''

Reflecting on her experience of switching in and out of groups, Tabitha said, ``I definitely think that could have been avoided. I don't think that anyone needs to have a bad experience on the first try in order to get it right. And I wish back in the day, when I talked to people, they had told me, `Oh, you're gonna go for work for this advisor, maybe you want to rethink that.'{''}  Hence, Tabitha believes that more open lines of communication with other graduate students might have informed her of this advisor's reputation and prevented her poor experience.  

%As a whole, the student experiences described in this section illustrate that departments must do more to . 

% In some cases, students discussed how they believed a diverse faculty and student body were indicative of healthy work environments.  While she was applying to graduate schools, Selena recalled wanting to know, ``am I going somewhere where there are other women, where there are people of color, where I am going to stand a chance of seeing real mentorship from someone who might get me?''  Above all else, this concern influenced where she chose to apply: ``If I go through your department page, and all I see are white men, I'm not clicking any further... There's no way that there's not a single woman or person of color who was good enough for your faculty. So that's telling me something about your department's attitude. And I'm just not going to look any further.''  Similarly, when looking for a new group, Pauline said, ``I looked through their group page to see the people who have been in there and who are currently in there, and see if there's any, for at least me, any kind of woman in the group or any kind of minority.'' The diversity of both the faculty as a whole and the individual groups therefore played a role in these students' search process.

%%%%%%%%%%%%%%%%%%%%%%%%%%%%%%%%%%%%%%%%%%%%%%

\subsubsection{\label{subsubsec:belonging}Navigating the search process impacts sense of belonging}

Our analysis revealed that the way students experience searching for a group is influential on their sense of belonging within the PhD program.  Of 20 interviewees, $N=4$ described how their perceived inability to get into a group left them feeling isolated and like they were not succeeding in the program.  On the other hand, $N=5$ discussed how successfully getting into a group made them feel a sense of connectedness and gave them motivation to continue working.   

When Lakshmi discovered that she might not be able to join her desired group due to lack of funding, she said ``it kind of scared me a lot at the end of the first semester. It was kind of my low point.''  This caused her to wonder whether she had made a mistake by choosing this graduate school: ``I was very low on confidence because of the other offer that I had rejected to come here. And that kind of made me feel like, did I do the correct thing? Did I make the right choice? So that was the point where I was like, `Okay, this is really not working out. What do I do?'{''}  Uncertainty over her ability to join a group led Lakshmi to doubt her commitment to graduate school overall.  Another PhD student, Brianna, described finding a research group as ``pretty rough.''  After four professors, whose research aligned with her interests, had told her they could not take on a new graduate student due to funding concerns, she said that ``Personally, I thought I was cursed. Like I slighted the department in some way, thinking that I had offended someone and I was being blacklisted from the research advisors.'' This comment illustrates how Brianna perceived this rejection as a personal failing. She continued, describing the process of looking for a group as ``just really depressing at points, because I was like, I guess I'm not meant to be here if I keep running into obstacles, that's the main thing for me. It's like I'm not supposed to be here... So that felt bad. And there's also like, comparing, I don't know why I'm not as far along as everybody else.'' This quote emphasizes how rejection from potential research groups influenced Brianna's sense of belonging and drove her to consider leaving the program (still, Brianna was eventually able to find a group and stay in the program; her story is detailed in Ref. \cite{verostek2023inequities}).

These experiences show how students who struggle with the process of finding a group can leave them feeling isolated and questioning their place in graduate school.  However, these feelings are avoidable. Brianna might never have felt that she was ``not supposed to be here'' if her department had been able to more effectively guide her into a research lab.  In fact, our data indicate that becoming part of a research group not only allows students to avoid these negative feelings of isolation, but can also significantly strengthen their sense of belonging in the program.  

Nathan joined a research group early in the first year of graduate school, which turned out to be a major factor in his second year decision to remain in his graduate program. Throughout his first year, Nathan's course grades were lower than he had been accustomed to in undergrad and he was feeling isolated within the department, saying ``the social side hasn't quite clicked here for me.''  This left Nathan questioning whether he was motivated and capable enough to pursue his degree in physics: ``Last semester was a whole mess with one of the classes, with stat mech... it just wasn't exciting me. And then the other class, the professor taught it really horribly. And so I was... that's when it muddled again with, is this a `me and physics' thing?''  However, being in a research group gave Nathan an outlet to discuss these feelings with his advisor, recalling ``That's also when I expressed to [my advisor] that I might not stay for the whole program, I might leave after a master's. And we talked about that for a while.''  Despite feeling disinterested in the course material, Nathan realized that the work he did in research ``felt really good and independent,'' which indicated to him that  ``I am still interested in physics, and there's still lots to do. That whole portion I still get very excited about... [my advisor] also wants me to stay in the group and keep doing the work, which is nice to hear.''  Thus, his advisor affirmed that Nathan was wanted in the group, and the other group members provided a community with whom he could talk about his struggles. Moreover, he found his research work more fulfilling than coursework, which contributed to his decision to stay in the program.  

Carmen, a first-generation college student, also had a difficult time adjusting to life in graduate school and considered leaving. However, joining a research group motivated them to stay in the program.  Carmen commented that upon entering graduate school they felt ``like both a fish out of water and a very small ant in a very large world.'' They continued, saying ``Grad school has not been a fun roller coaster ride. There have been other negative experiences within the department that have impacted me, to the point where I actually was considering dropping, like mastering out of the program.''  In particular, Carmen described a ``struggle with impostor syndrome, mainly being like, do I belong here? But also do I want to be here? I also felt like I lacked a sense of community. And so most of my thought process first year and parts of this second year has been, how can I make this feel like home?''  Yet Carmen remained in the program, directly citing their involvement with a research group as the reason.  They described how ``When it comes to my current research group, I felt a little bit more protected... being able to be a part of a research group made me feel like now I have people that I can talk to about my experiences, and who may give me some good advice on how to navigate these things. And also, I think I felt more comfortable with, like, failure.''  Group meetings in particular provided Carmen with a space where they could ``be given feedback on what I can do to, you know, improve things or even just like, have someone say, `good job.'{''}  Whereas Carmen never received the positive feedback they sought from other parts of the graduate program, their research group community provided constructive feedback that uplifted Carmen and gave them confidence to continue. 

The positive influence of joining a group on sense of belonging was evident in several other students' stories as well.  Taken together, these experiences indicate the major impact that joining a community of researchers can have on graduate students' careers.  Facilitating students' entry into a supportive group can give them a stronger feeling of community than they might be feeling in the rest of the program, and also help them avoid negative experiences associated with struggling to find a group.  Crucially, students with a higher sense of belonging are less likely to leave, which suggests matching students more efficiently with research groups may contribute to reduced attrition.

% In contrast to Nathan and Carmen, other students emphasized how difficulty finding a group left them feeling isolated and questioning their place in graduate school.  For instance, 

%%%%%%%%%%%%%%%%%%%%%%%%%%%%%%%%%%%%%%%%%%%%%%

\subsection{\label{subsec:difficulties}Difficulties finding a group}

\subsubsection{\label{subsubsec:unprepared}Students feel unprepared to make such a significant decision}

Despite the importance of finding a research group on both a personal level for students and a programmatic level for departments, many students reported that they felt unprepared to make such a significant decision.  Although all interviewees had completed four-year physics degrees and most participated in undergraduate research activities, $N=11$ students believed that they did not have adequate knowledge to understand what professors' research entailed, or did not know that certain research areas existed at all. Four students also reported feeling that they lacked expectations for what relational dynamics in a graduate school lab should look like, which made it difficult for them to evaluate a group's culture. 

Summarizing his academic and research career, Nathan reflected that he often felt unprepared to decide what topics to pursue, saying, ``I have always felt that my education lags behind the decision-making timeframe. I just feel like whenever I've needed to make a decision of leaning in a direction in terms of subfield, theory versus experiment, high energy versus condensed matter versus this other thing. Then some field within high energy. I feel like I keep needing to not necessarily make a hard decision, but start leaning in a direction before I know enough to even begin to make that decision.'' Nathan believed that his knowledge was insufficient for him to feel comfortable making decisions about his future research. Similarly, Blake reflected that upon arriving to graduate school and beginning to evaluate the research going on in the department, ``I think my main concerns were trying to read through papers and actually understand what was happening to a degree where I could decide whether or not it was interesting. Like, I can't figure out what's going on in this paper, so how am I going to figure out whether or not I want to join this group if I can't understand what they're doing at all?'' 

Students shared several of the strategies they used to grapple with these challenges. Similar to Blake, Olivia recalled feeling ``a bit blind'' in her group search because ``If you just read someone's thing about their research, it's very hard to understand what they're actually doing.'' To cope with this issue, she said ``I would just know of one concept that I thought was cool. And so I thought like, `topological insulators' was cool. Stuff like that. So I would look at things and if I saw that phrase, I'd be like, that's good.'' Matias also reported that much of his strategy for evaluating research groups came down to searching for ``buzzwords that I was really drawn to'' such as ``superconductivity, quantum Hall effect...''  However, both Olivia and Matias acknowledged that this strategy had limited effectiveness in helping them figure out whether they would enjoy a group's research on a day-to-day basis.  Another student, Selena, lamented that coming from a small school did not give her the opportunity to participate in a diverse range of astronomy research activities.  She described having to make ``pretty surface level decisions'' based on what she could ascertain about research from professors' websites.  ``I was like, alright, I guess I'll look at theory stuff... I don't know what's up with observation and data reduction and all of that.''  Being unfamiliar with those topics, Selena chose to focus on programs offering theory.   

Another reason students felt unprepared was they did not believe they had a good understanding of what working in a graduate research group should be like on an interpersonal level. Carmen discussed feeling like they did not have clear expectations for what a productive advisor-advisee relationship entails, and therefore turned to other students to help ``get a better understanding of how I should navigate my relationship with my advisor. Because, again, as a first-generation student, I don't know what that should look like. And I had no idea where to start... I'm still figuring out a lot of things. But a lot of it has been trial and error.''  Benjamin, a second year PhD student who switched research groups in the middle of his first year, similarly indicated that he felt unhappy in his first research group but assumed that being unhappy was simply inherent to the graduate experience: ``At the time, I didn't realize as much, maybe, the negatives... I felt like, you know, [my lab] is doing cutting edge science, it's gotta be hard, you're gonna have to suffer a little bit. That's what I thought at the time. I didn't necessarily think it was wrong that I was feeling anything negative about the research I was doing. I felt like that was just part of how it had to go.''  Thus, in the absence of expectations for a healthy group environment, students may assume that their unfulfilling research experience is normal.  

%As a whole, the excerpts shown in this section illustrate how  PhD students often felt ill-equipped to manage several important aspects of the group search process, including the evaluation of groups' research interests and knowing what to expect out of working in a graduate research group.

%%%%%%%%%%%%%%%%%%%%%%%%%%%%%%%%%%%%%%%%%%%%%%

\subsubsection{\label{subsubsec:structure}Wishing for more structure and guidance}

Feeling apprehension about their impending group search, students oftentimes expressed that they wished their department did a better job facilitating various aspects of the search process.  As summarized by Brianna, a first-generation graduate student, ``I was under the impression that I'd get a little bit more advising help, like how to navigate through grad school. And like I said, I really just had to figure it out on my own.''  Indeed, 17 of 20 interviewees made explicit comments describing ways in which they wished their department had provided more structure or guidance while they searched for a research group.  These statements most frequently revolved around wanting more efficient ways of connecting with faculty and students and wishing there were more easily accessible resources and explicit guidelines in place to direct them through the process.  

Six students reported feeling that their departments did not do an adequate job promoting communication between students and research groups. While reflecting on his experience navigating the group search process, Alex felt that departments overall ``leave the student to fend for themselves.  You figure it out.  We'll help you if you ask, but we're not really going to set up structures that promote your ability to efficiently meet with laboratories and students.''  Matias expressed a similar sentiment: ``I would say that the department still hasn't done a really good job at facilitating ways to learn more about groups here.''  He contended that if he had not earned a fellowship to work in a lab over the summer, ``I would still be very uncertain right now... I would know nothing of the culture of the group. And I don't like that.''  A lack of structured opportunities to meet faculty and older grad students represents a barrier to gathering information, and may be especially harmful for students who are less likely to seek out faculty independently.

Furthermore, $N=11$ interviewees said that they wished there had been additional resources from the department with advice on navigating the group search process.  Like Alex, Jack also used the phrase ``fend for yourself'' to emphasize his perception that the department did not provide much guidance, saying that ``Once you get here, you kind of have to fend for yourself to get a spot with one of the professors.''  Blake reported feeling ``a little confused about what my role was supposed to be in looking to join the groups.''  Blake recalled  not knowing when to reach out to professors, how in depth to read papers in preparation for meetings with professors, and what red flags to look out for in the group meetings they attended.  Luis, another first-generation graduate student, described feeling like there were ``no guidelines'' for finding a group, and remembered specifically thinking that ``the difficulty was like, not knowing where to reach out for help.'' 

In lieu of departmental guidance, students often described turning to peers for help.  Brianna reflected that in searching for a group, ``I think I made the mistake of trying to place my trust into the professors, when I probably should have been trying to make connections with other grad students, because they know what's going on.''  In this comment, Brianna suggests that the informal network of peers was more helpful than official departmental contacts. While peers undoubtedly provide valuable insight into navigating the doctoral process, relying on these informal networks to disseminate important information to graduate students opens the way for systemic inequities in access to information.  This effect is likely exacerbated for first-generation graduate students like Brianna and Luis, who both described feeling unsure of where to go or who to talk to in order to get help.  In contrast, two students noted that having a structured network of faculty and peers facilitated by the APS Bridge program was beneficial for navigating problems they encountered in graduate school.  As Pauline said, ``Bridge has this thing where every Friday we'll meet up with the other Bridge students and their PIs and have like, topics on different things like work-study balance, maybe how to prepare for grad school, dealing with impostor syndrome.'' 

%These results, combined with the results of the previous section, depict a system that graduate students feel simultaneously ill-equipped and unsupported to navigate.

%%%%%%%%%%%%%%%%%%%%%%%%%%%%%%%%%%%%%%%%%%%%%%

\subsubsection{\label{subsubsec:coursework}Coursework and research are perceived as being in conflict}

Classes are undoubtedly the most familiar and structured aspect of the first year of graduate school, as nearly every physics graduate student does coursework their first year.  On the other hand, research comprises the majority of the graduate school experience, which would seemingly make the process of joining a group a top priority for first year students.  Although the purpose of graduate coursework is ostensibly to prepare students for research, interviewees reported that emphasizing coursework actually hurt their ability to figure out what research they wanted to do.  $N=10$ interviewees reported feeling this tension between their research and coursework. 

Some of these students $N=5$ felt unclear on how to best divide their time between searching for a group and doing coursework.  Brianna recalled being unclear as to ``whether or not I had to find [an advisor] right away, as opposed to like, waiting out and focusing on trying to pass the courses first. Because I feel like doing all of that was a little bit stressful.''  Meanwhile, Selena intended to try out research with a prospective advisor during her first year, but was worried about balancing research with coursework.  She said, ``I wasn't clear, since I'm still taking classes, I was a little concerned going in what my expected output would be in terms of research versus classes. Because I know there's some advisors in the department who are like, classes don't matter. So I expect you to not prioritize them in favor of research. And then some who are very much like your grades are still gonna matter.''  For both Brianna and Selena, unclear expectations from the department and professors resulted in anxiety over how they should allocate their time.

Six students also indicated that doing well in graduate coursework required such a significant time commitment that it limited their capacity to explore research groups.  Benjamin explained that although he wanted to find out more about different groups' research by reading their papers, ``How much are you really going to get out of reading their very technical, complicated papers? Some people better than others. At the time for me, when I was trying to figure all this out I was in the middle of classes. I didn't want to read a bunch of papers to try to understand if this was interesting.''  He also commented that he wished he could have tried out different research groups during his first year, but feels like that is infeasible while taking classes: ``It would be nice if you could kind of hop around in the first year, even though you don't have any time... Like how can people get more experience without making that first year even more stressful? It's hard to remember how stressful it was, but that was the hardest year of my life. Without a doubt.''  Dev worried that he was falling behind in his group search due to this lack of time.  He said, ``It's most stressful because I cannot really devote a lot of time to [talking to professors] at this stage, given the courses, given the TA-ship. So in the back of my head, I always feel, okay, I should have done this today. But throughout the day there was no way I could have done it. And given that, it's a bit stressful.'' 

%Thus, the results of this and the previous section demonstrate that students often feel they do not get the support they want, and that the coursework they are required to do in their first year can actively hinder their group search.

%%%%%%%%%%%%%%%%%%%%%%%%%%%%%%%%%%%%%%%%%%%%%%

\section{\label{sec:Discussion}Discussion and Future Work}

\subsection{Departmental changes to support student retention and satisfaction}
% how better structures could help retention and satisfaction

% Departmental practices for supporting graduate students fundamentally determine who is allowed to shape the future of physics research.  It is therefore essential to critically examine how graduate programs currently operate in order to improve student outcomes and assure that the path to a physics PhD is inclusive and equitable.  In this paper, we investigated a crucial aspect of the graduate experience: finding a research group.  To do so, we interviewed physics graduate students to better understand their attitudes toward searching for a group, as well as how that process interacts with other elements of the graduate experience.  

% 1) Results show how this might be a lever to improve retention.  Sense of belonging was lower for students who couldn't find a group, higher for people in group.

Students characterized finding a research group as a significant event that impacts their persistence in the program as well as their careers beyond the PhD.  However, our analysis also revealed that it is a process many students felt unprepared to navigate and believed was underemphasized by their departments.  This aligns with previous research indicating that advisor selection processes in graduate physics programs tend to be poorly formalized \cite{artiles2023doctoral}; instead, classes are the most familiar and structured aspect of the first year of graduate school. Better facilitating the process by which physics graduate students find a research group may offer a variety of benefits for physics departments and students, including higher retention (particularly for traditionally underrepresented students) and improving students' overall satisfaction in the program.

Prior research has shown that fostering students' sense of belonging is one way to positively influence persistence in graduate school, especially among traditionally underrepresented students \cite{o2017sense, miller2020values, aip2020teamup}. Our results suggest that reforming the process by which students join research groups could provide a means for departments to improve students' sense of belonging, thereby supporting overall retention.  We observed that students already embedded in a research group described feeling more productive and part of a supportive community, whereas students who struggled to find a group felt isolated from their peers and in doubt about their ability to continue in their program.  If departments were able to guide students into productive research mentorships more efficiently during the first year of the program, more students could experience the benefits of doing research while avoiding the negative impacts of a difficult group search.  The fact that this intervention would come early in students' graduate careers may be particularly impactful, as physics PhD students are most likely to leave their programs during the first two years of study \cite{verostek2021analyzing}.

One idea to benefit students in the near-term is the development of an individual development plan (IDP) tool specifically for physics graduate students.  IDPs are tools designed to help students explicate their career goals and describe in detail how they plan to meet them \cite{bosch2013building, vanderford2018use, tsai2018optimizing}.  Hence, IDPs serve as a way to help students know what they should be doing at each step of their doctoral process in order to achieve their goals.  In 2012, an IDP was developed for use by PhD students across STEM \cite{hobin2014putting}. Recently, a tool specifically designed for students in the chemical sciences was also developed \cite{idp2023}.  If physics departments were to formalize the use of a similar tool upon entry as part of a student's general advising, it could help provide structure for those struggling to figure out how to best allocate their time during the first year.  A tool designed specifically for physics graduate students should explicitly emphasize exploring research options and joining a research group, including steps that students need to take in order to join a group and when those activities should take place.

Focusing more effort on helping students match with a research group may ultimately lead to higher satisfaction for both faculty and students in their advising relationships.  Faculty want to recruit graduate students who are passionate about the research they are doing.  However, our results showed that students sometimes felt unaware of what research was available in their departments, or unprepared to decide if research seemed interesting.  Research seminars, the most common method among physics departments for informing students about available research \cite{artiles2023doctoral}, are beneficial but clearly insufficient for providing students with the information they desire.  By comparison, increasing exposure to available research was one major benefit of the formalized rotation system used in the context of biology \cite{maher2019doctoral}. If physics students are unaware that a faculty's research area exists at all, they may miss the chance to take part in research that they are passionate about.  Physics departments should therefore consider implementing more formalized ways for students to experience working in different research groups, such as a rotation system.  Students who are doing research that aligns with their interests and career goals may enjoy their work more, and potentially result in higher productivity and reduced time to degree completion \cite{verostek2021time}.  In the words of one of our interviewees, Rose, ``We definitely have to get better with the overall pairing of students and PI, of students and research topics. Because that's how we get the best science: when people are in the right environment, in their element... We just have to have the right setting, the right group, the right everything, and we can get a lot more science done.''

% 2) better structure would have outsized impact on DEI efforts. SoB important for underrep students.  But also becasue we've shown how this aspect of doctoral process is susceptible to systemic inequities. Evidence to suggest this is the fact this is part of the hidde curriculum/ lots of norms and expectations about where to get help, what to spend time on, how to interact with faculty and grad students, what to look for, aspects often called the hidden curriculum grad students probably help a lot but this is insufficient and harms minority students in particular; thus not elucidating it contributes to ysstemic inequities, evidence that exposing hidden curriculum can help

%Students whose backgrounds afforded them prior familiarity with those norms were implicitly privileged, as they could navigate the group search process more efficiently.

% Yet our results demonstrate that finding a group can give students greater belonging and self-efficacy, and more students might be able to experience those benefits if they were able to become immersed in research groups more expeditiously.  Thus, adding structure and guidance to students' group search could have positive impacts on retention.

Finally, our results illustrated how students' search for a research group plays a role in perpetuating systematic inequities across race and gender in physics graduate education.  Norms surrounding where to reach out for help, how to best communicate with faculty and graduate students, and how much time to devote time to coursework were all important things for students to understand during their group search, but were seldom made explicit.  This collection of unwritten rules and norms is often referred to as the ``hidden curriculum'' \cite{snyder1970hidden, acker2002hidden, calarco2020field}, and can play a major role in reproducing inequity for underrepresented groups in physics graduate education \cite{apple2012education, bourdieu1973cultural, anyon1980social,gardner2011those, posselt2014toward}.

For instance, the lack of structured opportunities to meet faculty and older graduate students systematically disadvantages the 25\% of physics graduate students who identify as first-generation college students, as this group is less likely to reach out for help on their own \cite{ekmekcioglu2023navigating, nsf2022earned}.  Prior research affirms that having students rely exclusively on such informal lines of communication hinders first-generation doctoral students' ability to access and benefit from important information \cite{gardner2011those}.  Indeed, Brianna, Carmen, and Luis specifically cited their status as first-generation graduate students as putting them at a disadvantage relative to their peers.  However, this burden is not equally shared across demographics and disproportionately affects students of color. According to the Survey of Earned Doctorates, in 2022 46\% of African American and 44\% of Latino PhD earners identified as first-generation, while just 22\% of White and 25\% of Asian students did so.  Students new to graduate school cannot be expected to know all the ins and outs of the doctoral education process, and are more likely to succeed when given proper guidance and support \cite{dunham2012just}.  

Difficulties connecting with peers also produced disproportionately adverse consequences for women and non-binary students.  Tabitha and Pauline did not realize the importance of talking to peers about prospective groups (an unstated norm of the search process), which left them unaware of their respective advisors' disciplinary histories.  Lacking guidance from their departments or peers, both Tabitha and Pauline joined groups with a high likelihood of a negative mentoring relationship.    

%In fact, Pauline expressed surprise that the department allowed a faculty member with a disciplinary history to maintain his research group: ``It kind of made me a bit jaded to see that the PI of my former research group was still there, despite some serious allegations about him.''

Providing structures to better guide students searching for a group may therefore be particularly impactful for diversifying physics graduate education.  To this end, the American Physical Society Bridge Program offers several ``Effective Practices'' for departments to begin helping students find research groups \cite{aps_bridge}.  For instance, Bridge guidelines argue for integrating students into research group activities early in their graduate careers, which our results indicate would help students in build a stronger sense of belonging and limit attrition or thoughts thereof.  Furthermore, the guidelines urge departments to provide more opportunities for students to better understand what research is available to them, perhaps through mandatory research seminars led by PIs or senior graduate students.  This recommendation aligns with our finding that some students felt unprepared to choose a group because they were unaware of what other kinds of research existed in the department.  Another Bridge effective practice suggests that departments sponsor events for incoming graduate students to take lab tours with upper-level graduate students, aligning with our finding that students wanted more structured ways to meet graduate students in prospective groups.  One implementation of this might be for departments to devote one full day per semester to social and professional development, during which faculty and senior graduate students are made available to meet with first-year students at several times throughout the day without a pre-arranged meeting.  Departments could provide a template of topics for discussion, such as group norms, expectations, and day-to-day work in the lab.

\begin{figure*}[t]
\centering
\includegraphics[]{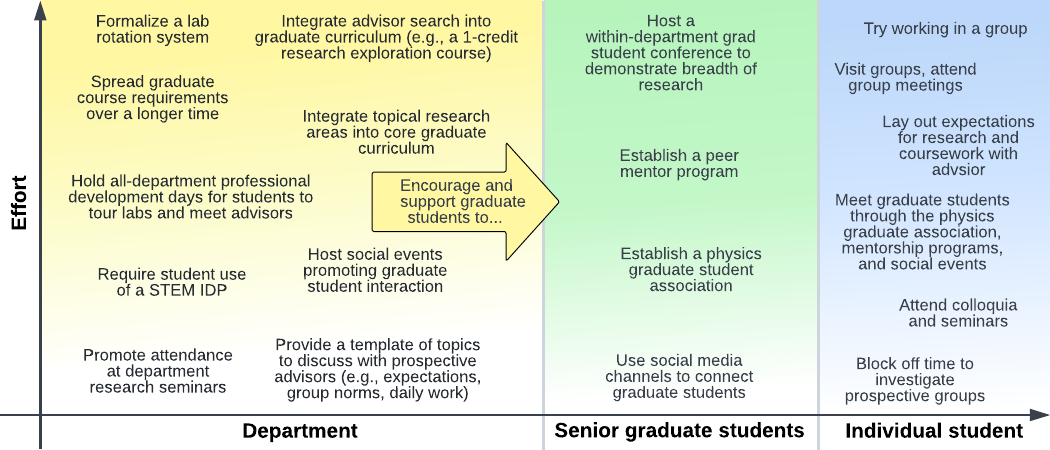}
\caption{\label{fig:discussion} A summary of suggestions to address challenges physics PhD students faced while searching for a research group.  Items are sorted by approximate level of effort required by the individual stakeholder to enact, relative to the other items in their list.  For instance, departments could adopt a policy to have first-year students complete an existing STEM-IDP with significantly less effort than enacting a formal research rotation program.}
\end{figure*}

\subsection{Helping students navigate the tension between coursework and research}

Another area in which departments can play a significant role is helping graduate students navigate the tension they perceive between coursework and research.  For students feeling unsure of how to navigate graduate school, as is often the case for first-generation graduate students, the existing structure sends the message that focusing on classes should be their top priority, perhaps at the expense of less formalized responsibilities like finding a group.  Indeed, some students were unsure of how to best divide their time between searching for a group and tackling coursework.  APS Bridge recommendations also recognize that students may struggle with balancing these responsibilities, and suggest that students who are experiencing difficulties passing classes should pare back expectations of research in order to make sure have time to dedicate to their coursework.  Presently, this recommendation likely aids retention since physics programs require students to pass classes in order to maintain good standing.

However, our results suggest that departments should reconsider whether it is desirable for students to focus so much on coursework during their first year.  We observed that students experienced more psychosocial benefits from being embedded in a research lab, \textit{not} from being in classes. Moreover, several students reported specifically feeling isolated \textit{due to} their coursework, and it was only because of their research groups that they felt part of a community. In fact, the time constraints associated with taking on a full course load during the first year actually hurt some students' ability to figure out what research they wanted to do, as it limited their ability to explore potential research interests.  Thus, we suggest that programs should consider ways of explicitly integrating students' advisor search into the graduate curriculum.  Combining these features of the first year experience would help alleviate students' perception that they have to sacrifice time looking for a group in order to focus on classes.  

% acknowledge they recognize this issue; since successful completion is required, advice is to reduce research expectations in order to make sure they can stay in their program.  but raises question for department as to whether they really want this to be the case. students feel their coursework and research identiies. faculty probably aslo have this. coursework associated with being a real physicist. so is research. but theyre divergent.

Furthermore, considering reforms to more systematically guide students into their research careers during the first year would also more closely align with students' first-year priorities. Our results show that finding the right research group is something that first-year graduate students highly value. On the other hand, our results suggest that physics graduate students do not view their coursework as particularly beneficial to their long-term goals.  This aligns with prior research by \citeauthor{busby2021program} in the context of chemistry doctoral programs, which showed PhD students tend not to value classes unless they relate directly to their research \cite{busby2021program}. This finding comes in spite of fact that the core chemistry curriculum is constructed to align with professional research areas (e.g., biochemistry, organic, inorganic, physical, analytical) \cite{acs2023guidelines}.  Since much of the core physics graduate curriculum is disconnected from modern areas of research, physics graduate students may value coursework even less than chemistry students.  Specialized courses that highlight technical skills or areas of faculty expertise might help students develop their research interests, but students are more likely to take these after completing the core courses or after they have joined a group. 

Revising course content to highlight available research in the department and allow students to explore potential areas of interest would more closely align with the goal of a curriculum that prepares students for research.  To assuage time constraints, courses could also be spread more evenly throughout the PhD timeline to give students more time for research and interest exploration during their first two years.  Professional development for instructors of graduate classes to enhance their pedagogical skills could help students learn course content more efficiently, and allow them to dedicate more time to thinking about their future research.

% Policy changes in physics programs across the country suggest that departments too are .  For instance, departments have begun replacing traditional qualifying exams with more research oriented approaches \cite{liera2023rethinking, aps2008graduate}.  In evaluating whether doctoral students are 

% As students in our data noted, some faculty also believe that ``classes don't matter.''

%Implementing the bridge recommendations is a step in the right direction but there may be additional factors that need to be considered.  
%Recs are conscientious of managable improvements palatable to wide variety of departments and faculty and so may push less hard for larger sturcutral changes

\subsection{Advice for incoming and current graduate students}

%For the first year student - blocking off some time for this.  Colloquia, lab tours, reading papers, meeting with graduate students, group meetings, talking to faculty.  even though there's no grade. 

% in lieu of formal advice, can visit grad students. cultivating relationships with other graduate students esp with those in senior years isn't just a social activity but a professional develppment tool.  graduate students one of best sources of help/source of info. observing norms/expectations. joining grad student organizations and getting into mentorship programs. some specific questions

Although we believe departments must improve how they guide students through their group search, we recognize that significant programmatic reform is a slow process.  We therefore also offer several suggestions for how senior PhD students can exercise agency and help to support new incoming students.  Upper-level graduate students can play an important role in helping new students to better understand what research opportunities are available and to more efficiently network with faculty and peers in prospective groups.  Moreover, senior graduate students are important stakeholders in the overall endeavor to improve physics graduate education, and their potential to affect positive change should be recognized.  We wish to emphasize however that senior graduate students are not independently responsible for implementing the suggestions outlined in this section; the department must still play an integral role in enabling them to happen by providing support and funding along the way.  

One promising option for helping first year graduate students navigate their group search is establishing a peer mentorship program.  Previous research has demonstrated that peer mentoring programs help facilitate the spread of tacit knowledge and expectations among new graduate students \cite{nokkala2022multidisciplinary, brown1999mentoring}.  Thus, with departmental support, establishing a readily accessible network of senior graduate students could help new students gain a better understanding of what groups are available and what they do.  This could include bringing students on lab tours to show them what day-to-day research looks like, or helping them set up meetings with prospective advisors.  It would also facilitate sharing knowledge about a group's culture and inclusivity, which our results demonstrated may be vitally important for some students.  

Mentoring programs could be facilitated by a physics graduate student association, which may also serve a role in sponsoring other events aimed at helping new students find groups. With department funding, senior graduate students could host a small within-department ``conference'' to give first-year students a better idea of what research is happening.  A poster session would let new students network with older peers and see what projects might fit their interests, as well as help to build community in the department.  Social media platforms may also play a role in allowing students to freely disseminate information to one another.  Regardless of structure, events should focus on ways to efficiently transmit information to all new graduate students equally, without them having to seek it out.  Students cannot ask questions about research if they do not know it exists.       

We also provide several suggestions for how new PhD students might be able to better navigate the existing system. In light of our findings, first year students would be well-served by blocking off time in their schedule to dedicate to researching prospective groups.  Even though it might not be graded, this time is just as important as a class.  For students who already have an advisor in mind upon entering graduate school, the time could be spent going to the lab, attending group meetings, shadowing a more senior student, and perhaps beginning to do some aspects of research work.  Talking to a prospective advisor about expectations for work is also important \cite{branchaw2020entering, barnes2009nature}.  Students who are less sure of the research they want to do should attend colloquia and seminars as much as they are able. Students can also read papers to get a feeling for what kinds of work is going on in the department, although several interviewees reported how difficult it was for them to understand the content of faculty research papers.  Moreover, these resources seldom give a good indication of what it is like to work in the lab on a day-to-day basis.  Stopping by group meetings, doing lab tours, and discussing potential projects with faculty are better indicators of what the workday will be like.

In lieu of formal information systems, cultivating relationships with senior graduate students is particularly important. Meeting other graduate students is not merely a social activity; rather, it is a professional development tool that can help in a variety of ways, including networking and understanding departmental norms.  Research has shown that the ``student grapevine'' is a critical means for new graduate students to gain access to important information \cite{gardner2007heard}.  Joining physics graduate student organizations, joining a mentorship program, and attending departmental events are all ways to meet other graduate students. Students might know of labs that are seeking new members and can help to meet students in those groups.  Talking to graduate students in prospective labs about their experience is crucial for understanding what their day-to-day research is like, as well as whether the lab seems like a healthy working environment.

\subsection{Open questions and future research}

Despite the recommendations given above, they remain incomplete.  Although this paper offers preliminary glimpses into how departments might improve the group search process for students, the primary goal of this work is to call attention to the importance of studying this phenomenon in greater detail.  Significantly more work is needed to decide how to best formalize the search process in order to provide the most benefits to students.  As our results have shown, this is a critical endeavor that has the potential to improve the entire graduate school experience for students.  This paper therefore serves as a starting point for a broad new area of research into physics graduate education that should be explored. 

No research to date has systematically characterized the process by which physics graduate students find a research group.  It remains unclear exactly what types of things physics graduate students are looking for when they search for a group, what actions they take to gather information, and what gets in the way of gathering that information.  It is also unknown at what point during their academic careers students begin prioritizing their group search (e.g., undergraduate senior year, first year of graduate school).  Determining what specific aspects of this process are the most difficult, as well as which are most helpful, would allow for more targeted interventions to support students at different stages of their academic careers.  Assessing the impact of any new programmatic elements will also be essential.

Future research into the group search process must also attend to how it is experienced across demographic groups and within different institutional contexts, as our study was limited in these regards.  Although our sample was diverse in many ways, our protocol did not specifically probe how students felt their identities impacted their group search.  The differences across gender that we discussed here arose naturally, but we see suggestions that other differences exist as well.  For instance, the three international students in our sample reported several unique challenges associated with applying to graduate schools and communicating with prospective research advisors.  Difficulties included understanding the differences in application processes between the United States and Europe, and navigating cultural norms surrounding what questions are appropriate to ask faculty.  A larger sample would be required to make more substantive claims about how experiences across demographics may vary.  We also did not probe how differences in how institutional size, location, and culture impact students' group search.  How do these characteristics play a role in the systems and structures available to guide students into research?  Gaining greater insight into how students value different aspects of the graduate school experience will also help us give them more holistic guidance and support.  

Furthermore, although our data included several students who considered leaving their programs,  we did not collect data from students who left their graduate programs and are therefore missing these important perspectives on the group search process.  Additionally, students whose group search experiences were particularly difficult may have been less likely to want to share those experiences in an interview setting.  Our sample may therefore be biased toward students whose experiences looking for a research group were more positive than the average.  International students may have also been hesitant to participate in an interview to discuss issues finding a research group, since their position in the US is dependent on good standing in their departments.  

Lastly, although they serve as useful recommendations for all students, we note that APS Bridge recommendations are not tied to specific research findings, and therefore are likely incomplete.  Moreover, they were designed specifically to ``increase the number of physics PhDs awarded to underrepresented minority students, identified as Black, Latinx, and Indigenous'' \cite{aps_bridge}.  Thus, they may not take into account the full range of differences across gender, international status, and other student identities.  For example, we observed that women and non-binary students were more acutely aware that joining an inclusive group would be important for them to persist through the PhD.  For some, screening groups for indicators of an inclusive environment led to limited research opportunities compared male students.  This is obviously a serious issue that must be addressed, but how to best bring about such change is unclear. 

This analysis has provided crucial insights into students' attitudes toward searching for a research group, as well as how the process interacts with other programmatic elements of the graduate experience.  We hope that our work serves to inspire other researchers to more critically examine this formative graduate student experience in order to to make PhD programs more inclusive, supportive, and productive for all students.

\begin{acknowledgements}
We thank the graduate students who participated in this study. We hope their stories contribute to ongoing improvements to doctoral education in physics and beyond.  We also wish to acknowledge Christian Cammarota for help with inter-rater reliability.  This work is supported by NSF Award HRD-1834516.
\end{acknowledgements}

\clearpage

\bibliography{main.bib}% Produces the bibliography via BibTeX.

\end{document}